\documentclass[10pt,a4paper]{article}

\usepackage[utf8]{inputenc}
\usepackage{amsmath,amssymb,pdfpages,geometry,nicematrix,graphbox,graphicx,placeins, mathtools, tikz-cd, upgreek, booktabs, multirow, siunitx} 
\usepackage[hidelinks]{hyperref}
\usepackage{appendix}
\usepackage{float}
\usepackage{authblk}

\usepackage{microtype}

\usepackage[labelfont=bf]{caption}
\usepackage{pgfplotstable}
\pgfplotsset{compat=1.17}

\usepackage[style=numeric-comp,sorting=none,autocite=superscript,
            doi=true,isbn=false,url=false,eprint=false,
            maxbibnames=99,giveninits=true]{biblatex}
\renewcommand{\vec}[1]{\mathrm{\mathbf{#1}}}
\newcommand{\angstrom}{\textup{\AA}{}}

\begin{document}

\title{Bayesian determination of structural ensembles from single-particle X-ray scattering images}

\author{Steffen Schultze}
\author{Nicolai Kozlowski}
\author{Helmut Grubmüller}
\affil{Max Planck Institute for Multidisciplinary Sciences, \\Am Fassberg 11, Göttingen, 37077, Germany}

\date{\today}

\maketitle

\begin{abstract}
Single-particle X-ray scattering with free electron lasers records snapshots of individual molecules with femtosecond time resolution and is therefore ideally suited to study structural heterogeneity or molecular motions. Current approaches to determine structural ensembles rely on prior classification of the scattering images, which is error-prone and becomes infeasible for single macromolecules, whose images contain too few coherently scattered photons. Here, we present a Bayesian approach that samples weighted ensembles of electron densities directly from the posterior probability distribution given all scattering images, thereby avoiding classification altogether. We demonstrate our approach in the low signal-to-noise high heterogeneity regime using synthetic scattering data of a small and flexible protein, as well as on published experimental scattering images for the coliphage PR772 virus. 
\end{abstract}

\section{Introduction}
\begin{figure}
    \centering
    \includegraphics[width=0.45\textwidth,trim={1.0cm 1.0cm 1.0cm 1.0cm}]{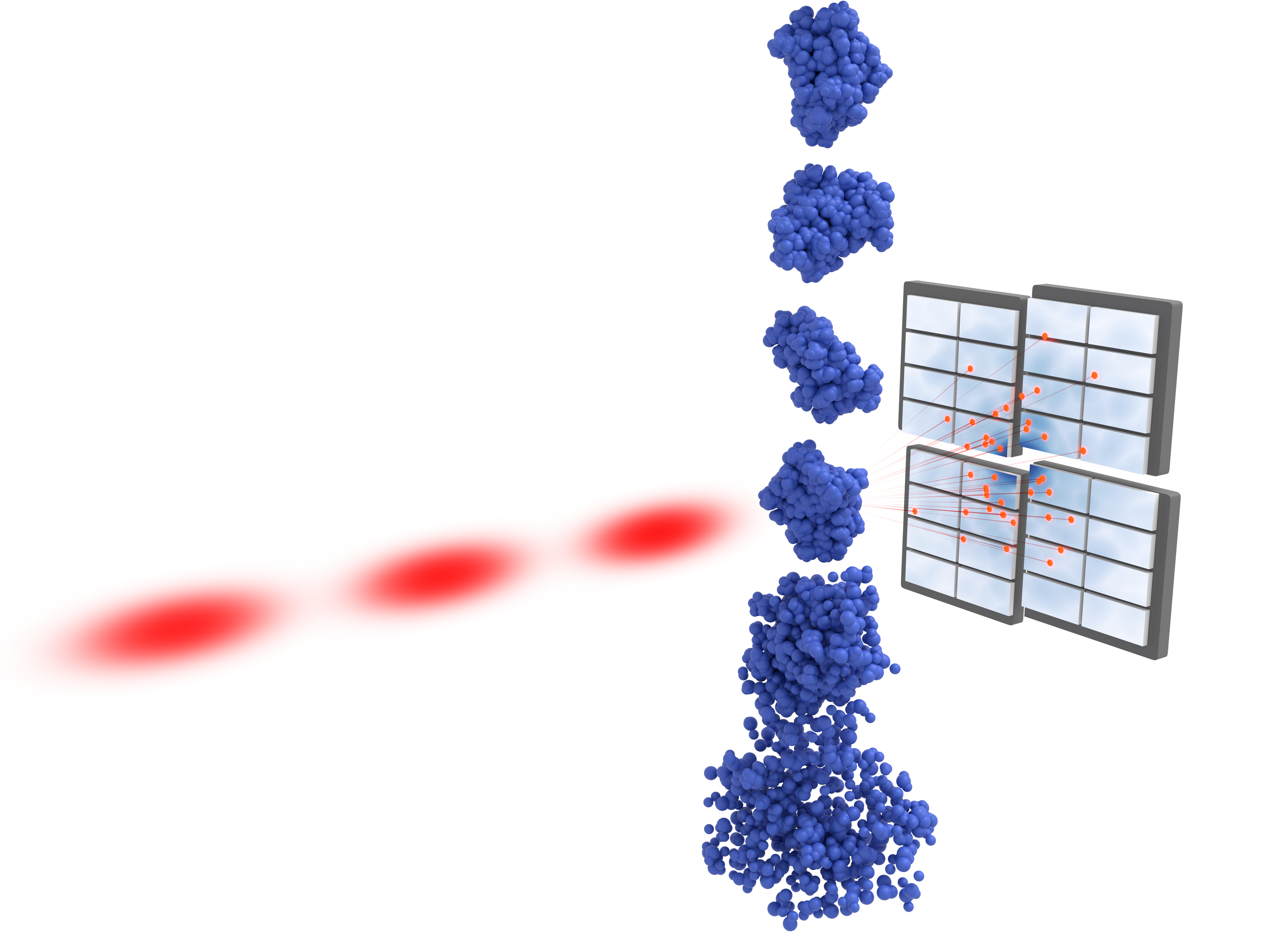}
    \caption{Single molecule scattering experiment. A stream of target particles is injected into the path of ultrashort, high-intensity X-ray pulses. For each scattering event, the scattered photons (red dots) are recorded as an image on a detector. Each image is thus a femtosecond snapshot of a single particle in an unknown, random orientation.}
    \label{fig: introexperiment}
\end{figure}

Single-particle X-ray scattering experiments using ultrashort X-ray free electron laser pulses (XFELs) allow for electron density determination approaching atomistic spatial and femtoseconds time resolution~\cite{hajdu_single-molecule_2000, huldt_diffraction_2003, gaffney_imaging_2007, miao_beyond_2015,chapman_x-ray_2019}.
In these experiments (Fig.~\ref{fig: introexperiment}), known as `diffraction before destruction'~\cite{chapman_diffraction_2014}, the sample particles are exposed to high-intensity few-femtosecond X-ray pulses, and the resulting scattering images are recorded on a detector. The short pulse duration not only ensures that the diffraction signal is collected before the sample is destroyed, so that it is essentially free of radiation damage~\cite{neutze_potential_2000}, but also gives each image femtosecond time resolution. Due to the latter, single-molecule XFEL scattering is ideal obtain not only averaged structures but to resolve ensemble information for highly dynamical targets like proteins or other biological samples, and holds the promise of time-resolved structure determination at high temporal and spatial resolution~\cite{hajdu_single-molecule_2000, huldt_diffraction_2003, gaffney_imaging_2007, miao_beyond_2015}.

However, obtaining structural information from XFEL scattering is challenging already for static particles with no structural heterogeneity, mainly due to unknown particle orientations and very low signal-to-noise ratios in the extreme Poisson regime~\cite{von_ardenne_structure_2018}. 
To address these challenges, a range of specialized analysis methods have been proposed, including, including orientation determination methods~\cite{shneerson_crystallography_2008, loh_reconstruction_2009, walczak_bayesian_2014, kassemeyer_optimal_2013, elser_three-dimensional_2011, tegze_atomic_2012, flamant_expansion-maximization-compression_2016, ayyer_dragonfly_2016}, manifold embedding~\cite{fung_structure_2009,schwander_symmetries_2012, giannakis_symmetries_2012, winter_enhancing_2016}, phase retrieval methods~\cite{elser_searching_2007, luke_relaxed_2004}, or photon correlation methods~\cite{saldin_structure_2009, saldin_reconstructing_2011, saldin_beyond_2010, saldin_new_2011, saldin_structure_2010, starodub_single-particle_2012, kurta_correlations_2017, donatelli_iterative_2015, von_ardenne_structure_2018}. 
These approaches have been successfully applied to a great variety of specimen, including nano-crystals~\cite{schlichting_serial_2015, chapman_structure_2017,oda_time-resolved_2021,tenboer_time-resolved_2014} and entire viruses~\cite{seibert_single_2011, ekeberg_three-dimensional_2015, hosseinizadeh_high-resolution_2014}.

While it has been proposed to perform these X-ray scattering experiments also on single macromolecules~\cite{neutze_potential_2000}, their relatively small size poses further challenges. The main obstacle is that they scatter much fewer photons per image than, for example, typical viruses ($10$ to $100$ as opposed to hundreds of thousands), rendering orientation determination highly error-prone or even impossible.

We have therefore recently developed a rigorous Bayesian treatment~\cite{schultze_bayesian_2024,schultze_structure_2025} that achieves electron density determination even at such extremely low photon counts. Rather than attempting to determine the particle orientation for each scattering image, it directly maximizes or samples from the Bayesian posterior probability distribution for the electron density given the whole set of images.

The next crucial step is to go from single average structure determination to resolving entire structural ensembles, which poses additional challenges. Chiefly, very much like in cryo-electron microscopy (cryo-EM)~\cite{scheres_relion_2012,punjani_cryosparc_2017,dashti_trajectories_2014, levy_cryodrgn-ai_2025}, conventional approaches~\cite{schwander_symmetries_2012,giannakis_symmetries_2012,winter_enhancing_2016,barty_cheetah_2014,andreasson_automated_2014,yoon_unsupervised_2011,jiao_predicted_2024} as well as recent approaches based on machine learning~\cite{assalauova_classification_2022,levy_cryodrgn-ai_2025} rely on a per-image assignment into conformational classes, and determining a structure for each class. This classification is challenging and error-prone, however, and may lead to biased ensembles~\cite{evans_counting_2026}. Similar to the orientation problem, these difficulties are exacerbated for single macromolecules, where the extremely low signal-to-noise ratios likely prohibit such classification altogether. For cryo-EM, a number of Bayesian methods have therefore been developed that avoid classification by working directly on the particle images~\cite{cossio_bayesian_2013,cossio_bioem_2017}, including ensemble reweighting~\cite{tang_ensemble_2023} and optimization~\cite{silva-sanchez_cryo-electron_2026}, free-energy profile inference~\cite{giraldo-barreto_bayesian_2021}, simulation-based inference~\cite{dingeldein_amortized_2025}, and structure-prediction-guided atomic model inference~\cite{maddipatla_foldem_2026}, typically relying on prior structural models such as atomic models or simulated conformations.

Here we develop and demonstrate a Bayesian approach for structural ensemble determination from single-particle X-ray scattering images that circumvents any need for classification, thereby producing a rigorous ensemble free of classification bias. Crucially, by sampling the full Bayesian posterior with Hamiltonian dynamics, our approach extracts structural ensembles even from extremely limited information, making it ideally suited for the sparse, noise-dominated scattering images obtained from single proteins. It also yields rigorous uncertainty estimates and requires no precomputed ensembles or prior structural models.

We demonstrate our approach on synthetic scattering images of a small, flexible protein, where the ground-truth is known, and on published experimental scattering images of the coliphage PR772 virus~\cite{reddy_coherent_2017}. For the latter, we recover a continuous conformational transition associated with the reorganization of the viral genome, consistent with a previous coordinate-based analysis~\cite{hosseinizadeh_conformational_2017}. Rather than being imposed from the outset, this conformational coordinate emerges directly from the ensemble, pointing to a natural route towards `molecular movies' from single-particle XFEL scattering, a long-standing goal of the field~\cite{ourmazd_cryo-em_2019,van_thor_advances_2019}.

\section{Results}
\subsection{Method overview}
Our approach considers the Bayesian posterior probability $P(\rho_{1..K},w_{1..K}\,|\,\mathcal I)$ of a weighted ensemble of $K$ electron densities $\rho_{1..K}$ with weights $w_{1..K}$, given the full set of observed scattering images $\mathcal I$. Rather than first assigning individual images to conformational classes and then reconstructing each class separately, the entire ensemble is thus inferred in a single step.
By Bayes' theorem, this posterior is, up to normalization, given by the likelihood of the images and the prior,
\begin{equation}
    P(\rho_{1..K},w_{1..K}\,|\,\mathcal I) \propto P(\mathcal I \,|\, \rho_{1..K},w_{1..K})\, P(\rho_{1..K},w_{1..K}).
\end{equation}
As the images are statistically independent and each image originates from one of the ensemble members, chosen with probability $w_k$, the likelihood decomposes into a product over all $N$ images of a weighted sum over the ensemble,
\begin{equation}
    P(\mathcal I \,|\, \rho_{1..K},w_{1..K}) = \prod_{j=1}^N \sum_{k=1}^K w_k\, P(\vec k_1^{(j)}, \dots, \vec k_{n_j}^{(j)} \,|\, \rho_k),
\end{equation}
where, for each scattering event $j = 1,\dots,N$, the corresponding image is described by the positions of the $n_j$ scattered photons on the detector, which are converted into scattering vectors $\vec k_1^{(j)}, \dotsc, \vec k_{n_j}^{(j)}$.

Crucially, the conformational class of each image is therefore marginalized rather than determined.
The single-image and single-density likelihood $P(\vec k_1, \dots, \vec k_n \,|\, \rho)$ describes the distribution of all possible scattering images for a given electron density $\rho$ and is determined by the physics of the scattering event in a similar approach as published previously~\cite{schultze_bayesian_2024}. In the same way, it marginalizes over the unknown orientation of the particle and accounts for background scattering and the detector geometry.
To sample from the posterior, each electron density $\rho_k$ is represented in real space as a superposition of identical Gaussian beads, and the prior combines a Dirichlet prior on the weights with generic, weakly informative priors on the bead positions. Bead positions and weights are then sampled jointly using thermostatted Hamiltonian dynamics~\cite{bussi_canonical_2007}. A detailed description of the likelihood, the structure representation and priors, and the sampling procedure is given in the Methods section.

We applied our approach to two complementary test cases. First, to assess whether ensembles can be recovered in the extreme Poisson regime expected for single proteins, we used synthetic images of a small, highly flexible protein with only about $30$ recorded photons each, for which the ground-truth ensemble is known. Second, to demonstrate applicability of our approach to real experimental data, we used published scattering images of the coliphage PR772 virus, whose structural heterogeneity has been characterized previously~\cite{reddy_coherent_2017,hosseinizadeh_conformational_2017}.

\subsection{Application to single-molecule scattering images}
\begin{figure}[p]
    \centering
    \includegraphics[width=\textwidth]{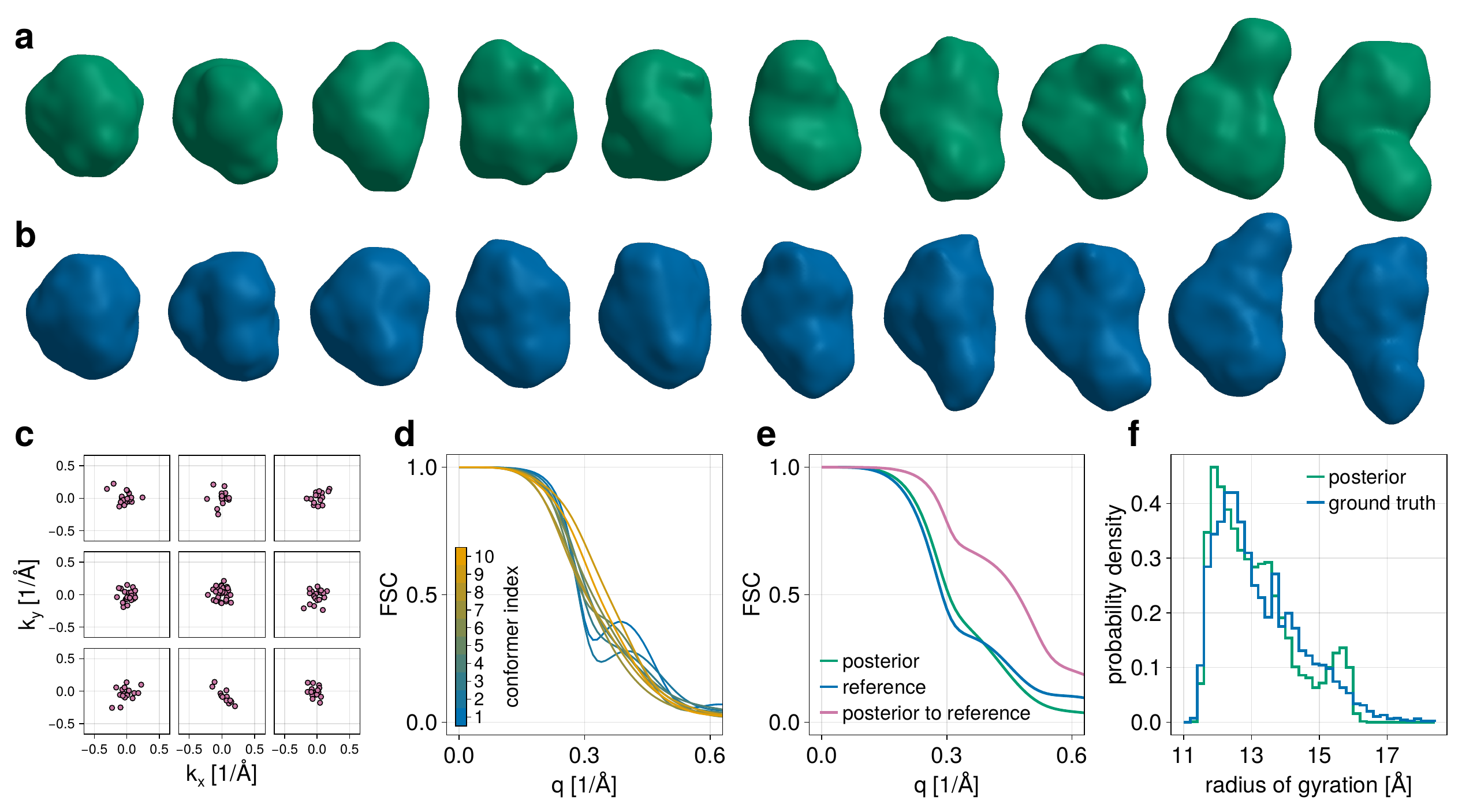}
    \caption{Ensemble reconstruction for the synthetic test case PDB~1GH9.
    \textbf{a}~Ten ensemble members determined from $50\,000$ simulated XFEL single-particle scattering images (about 30 photons per image), shown as electron densities averaged over posterior samples after rotational alignment and K-medoids sorting.
    \textbf{b}~Independently clustered and aligned ground truth ensemble, sorted by radius of gyration and smoothed by a Gaussian kernel $(\sigma = 2\,\angstrom)$ for visualization. 
    \textbf{c}~Exemplary simulated scattering images.
    \textbf{d}~Uncertainty within each posterior conformational class measured by average pairwise Fourier shell correlation (FSC) between sampled posterior densities.
    \textbf{e}~Average pairwise FSC for each ensemble member in posterior and reference ensembles, as well as FSC between average posterior and reference densities for each ensemble member.
    \textbf{f}~Radius-of-gyration distribution of the posterior and ground-truth ensembles. For the former, the densities from all sampled posterior ensembles were combined into one histogram.
    }
    \label{fig: 1gh9}
\end{figure}

As a test system for the single-molecule regime (Fig.~\ref{fig: 1gh9}), we chose the $8.3\,\mathrm{kDa}$ protein PDB 1GH9~\cite{berman_protein_2000,christendat_structural_2000}, whose particularly high structural heterogeneity makes it a challenging test case. As the ground truth, we used an ensemble from $100\times 10\,\mu\mathrm{s}$ molecular dynamics (MD) simulations and simulated $50\,000$ scattering images (as shown in Fig.~\ref{fig: 1gh9}c) from this ensemble, each from a randomly chosen trajectory frame. We here assumed negligible background scattering for simplicity, and normalized the intensity such that the resulting images contained a realistic average $30$ photons each~\cite{schultze_structure_2025}.

Using this set of images, we then sampled the posterior for an ensemble of ten independent structures with uniform weights, each described by $50$ Gaussian beads. 
Because the densities in the resulting posterior ensembles are randomly orientated and shuffled, they were then rotationally aligned and sorted using a K-medoids algorithm (see Methods).
Figure~\ref{fig: 1gh9}a shows the average electron density for each of the resulting aligned and sorted conformational classes. For comparison, a similar equal-weight K-medoids clustering was performed on the ground-truth MD-ensemble (Fig.~\ref{fig: 1gh9}b). Importantly, the conformational sorting of the sampled ensembles and the ground-truth were separate, merely the final averaged densities were aligned to each other for visualization. As can be seen, the ground-truth clustering and the reconstructed ensemble match well. 

To quantify the uncertainty of each ensemble member and to estimate a resolution, we calculated the average pairwise Fourier-shell-correlation (FSC) curves between the sampled posterior densities for each ensemble class (Fig.~\ref{fig: 1gh9}d). As can be seen, at a conservative cutoff of $0.5$ the posterior uncertainty corresponds to an average resolution of about $2\,\mathrm{nm}$. 
For comparison, we calculated the same average pairwise FSC curves for the reference ensemble, i.e., between the ground-truth structures within each class (Fig.~\ref{fig: 1gh9}e). The curves for posterior and reference are nearly identical, indicating that the spread within each posterior class reflects the structural variance within the corresponding reference class. Accordingly, the FSC between the average posterior and reference densities of each class is higher, corresponding to a resolution of about $ 2\pi / (0.45\,\angstrom^{-1}) \approx 1.5\,\mathrm{nm}$. An alternative, real-space uncertainty measure based on optimal transport yielded the same picture (Fig.~S1), with mean pairwise Wasserstein distances of about $3\,\angstrom$ both for posterior and reference, and of about $1\,\angstrom$ between the average densities. Therefore, we conclude that the resolution was here limited by the number of ensemble members.

To further assess if ensemble properties were recovered correctly, we compared the distributions for the radius of gyration obtained from posterior and ground truth (Fig.~\ref{fig: 1gh9}f). As can be seen, the two distributions agree well. Some deviations can be seen, however, for example at the large $r_g$ end. We attribute these differences to the finite posterior ensemble size of ten, which prohibits the posterior ensemble from covering phase-space areas of very low occupancy in the ground-truth distribution. 

\subsection{Application to a virus data set}
Having validated our approach on synthetic data, we next applied it to published experimental scattering images of a single coliphage PR772 virus~\cite{reddy_coherent_2017}. This data set has been analyzed using a variety of other methods~\cite{hosseinizadeh_conformational_2017,kurta_correlations_2017,lundholm_considerations_2018,assalauova_classification_2022,xian_variance-reduced_2022,schultze_bayesian_2024,shenoy_scalable_2025,chen_augmenting_2025,berberich_imaging_2026}, rendering it a particularly informative benchmark. In particular, the virus has been shown to exhibit considerable structural heterogeneity due to a reorganization of the genome inside the virus~\cite{hosseinizadeh_conformational_2017}.

\begin{figure}[p]
    \centering
    \includegraphics[width=1\linewidth]{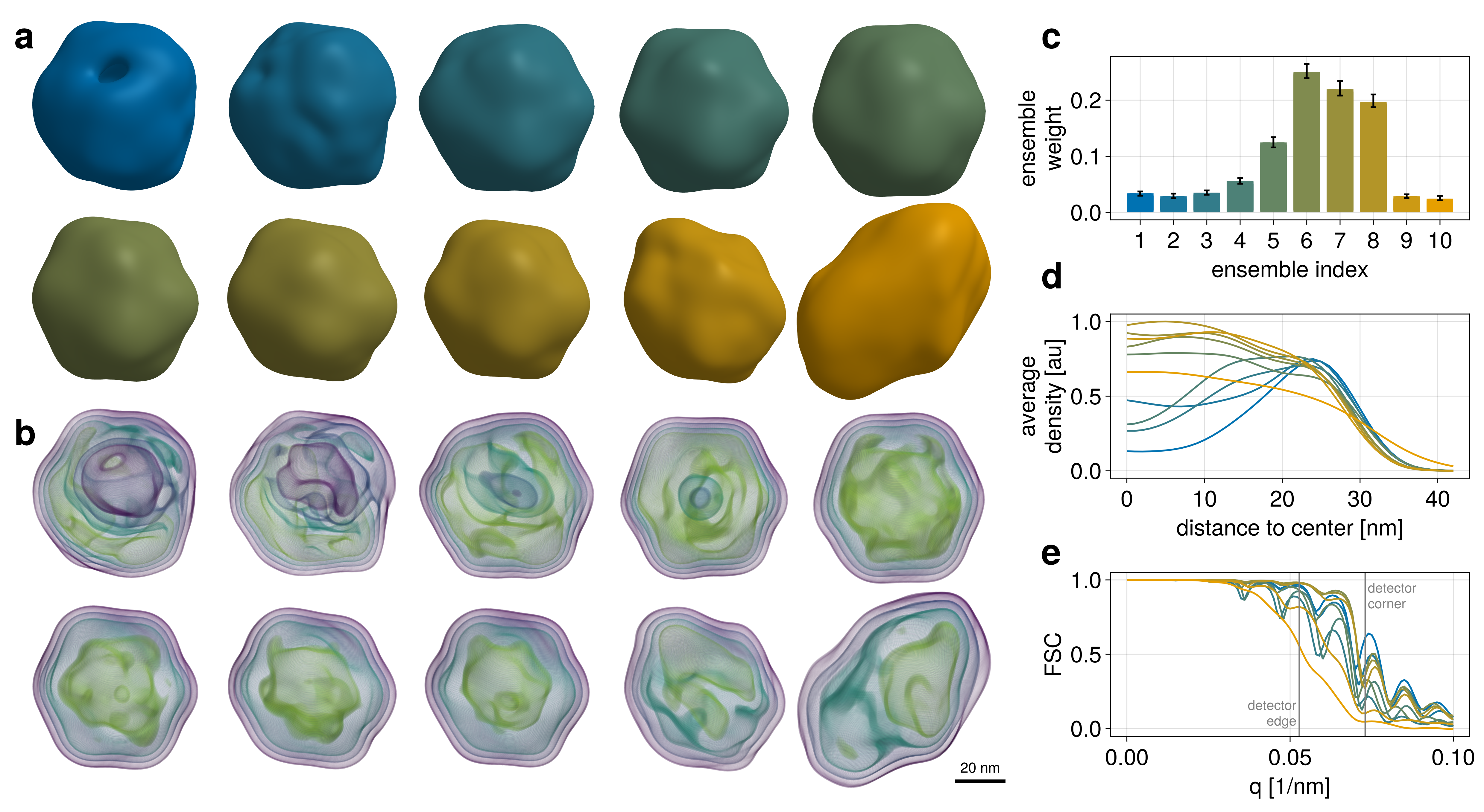}
    \caption{Conformational ensemble of coliphage PR772 determined from experimental scattering images. The ten ensemble members are ordered manually along the apparent conformational coordinate, from depleted interior (top left) to filled interior (bottom right), and are numbered row-wise; colors identify ensemble members consistently across all panels.
    \textbf{a}~Isosurface of each averaged ensemble member at $0.2$ of its maximum density.
    \textbf{b}~Nested semi-transparent isosurfaces of the same densities at equispaced levels from $0.2$ to $0.8$ of the maximum, revealing the internal density distribution.
    \textbf{c}~Average ensemble weight of each ensemble member; error bars indicate the minimum and maximum over posterior samples.
    \textbf{d}~Spherically averaged radial density profile of each ensemble member.
    \textbf{e}~Uncertainty within each ensemble member measured by average pairwise Fourier shell correlation (FSC) between aligned densities drawn from posterior samples of the same ensemble member. Vertical lines mark the momentum transfer $q$ at the detector edge and corner.} 
    \label{fig: coliphage}
\end{figure}

As before, we sampled a weighted posterior ensemble consisting of ten densities (Fig.~\ref{fig: coliphage}). Compared to the above 1GH9 test case, each density was described by a much larger number of $4000$ Gaussian beads. This number of beads is required to represent the mostly uniform interior of the virus with its relatively sharp outer boundary.
Further, these data are much more informative, making the posterior much harder to sample properly. Indeed, the photon counts of the full images are so large that slight inaccuracies in, for example, the modeled gain per pixel can have large effects on the resulting densities. To account for such unmodeled systematic errors, which the Poisson likelihood alone does not capture, we raised the likelihood of each image for each pose to the power of $0.1$ before marginalizing over poses (see Methods). Moreover, we added an ensemble-level Gaussian prior that couples the ensemble members to their mean, such that they deviate from a common structure only where supported by the data (Methods, Section~\ref{sec: structure representation}). This prior is also important for efficient sampling, as it pulls poorly fitting ensemble members towards the mean structure, where the likelihood provides informative gradients; otherwise, such members might receive vanishing gradients and remain trapped.

Figures~\ref{fig: coliphage}a and b show the resulting average posterior densities, Fig.~\ref{fig: coliphage}c their average weights. Sorted manually, they form a sequence that resembles snapshots of a single, continuous conformational transition. From left to right, this sequence shows the gradual filling of an off-center hole combined with decrease of the virus diameter, as can also be seen from the radially averaged densities (Fig.~\ref{fig: coliphage}d).
Two densities (Fig.~\ref{fig: coliphage}b, densities~9 and~10) do not fit into this sequence. Density~9 resembles a skewed version of the icosahedral density, whereas density~10 is of markedly larger size. We attribute density~10 not to an actual state of the virus but to images where the X-ray pulse simultaneously hit several particles, which are known to be present in the image set used here~\cite{reddy_coherent_2017}. This example illustrates a strength of our approach: because such outliers are absorbed by a separate ensemble member rather than distorting the remaining densities, these multiple hits do not need to be filtered out in advance.

In contrast to the previous test case, we here did not observe label switching or rotational drift. To assess convergence, we therefore repeated the sampling from a different random initialization (Fig.~S2). Both runs yielded qualitatively the same ensemble, but differed in finer internal features of individual densities and in the precise weights of individual members. We therefore consider the conformational sequence and the coarse weight distribution converged, but not smaller structural details or individual weights.

Our determined ensemble is consistent with an earlier analysis~\cite{hosseinizadeh_conformational_2017}, which identified a similar conformational coordinate through image classification, and the ensemble weights (Fig.~\ref{fig: coliphage}c) agree qualitatively with the distribution found there. The considerable deviation from icosahedral symmetry observed in some of the densities has likewise been reported before~\cite{kurta_correlations_2017}. The resolution is also on par with that of earlier work; indeed, average pairwise FSC curves (Fig.~\ref{fig: coliphage}e) yield about $9\,\mathrm{nm}$ for most ensemble members, which corresponds to the limit imposed by the detector geometry. 
We find no evidence in our ensemble for the rare conformer with a protruding tubular structure reported previously~\cite{hosseinizadeh_conformational_2017}.

\section{Discussion and Conclusion}
We here developed and assessed a rigorous Bayesian approach to determine electron density ensembles from XFEL single-particle scattering images. Rather than classifying individual images and reconstructing one structure per class, our approach samples the joint posterior of all ensemble members and their weights given the full data set, marginalizing over both the orientation and the conformational index of each image.
In contrast to related ensemble methods for cryo-EM~\cite{cossio_bayesian_2013, cossio_bioem_2017, tang_ensemble_2023, giraldo-barreto_bayesian_2021, dingeldein_amortized_2025, maddipatla_foldem_2026, evans_counting_2026, silva-sanchez_cryo-electron_2026}, which use an analogous ensemble likelihood function but typically rely on prior structural models such as atomic models or simulated conformations, our approach samples the full posterior using only generic priors, such as compactness and smoothness.

A key advantage of our approach is that it requires no conformational sorting of individual images. For single-molecule targets such as our 1GH9 test case, this is essential, as the conformational index of each individual image is underdetermined. Any assignment of single images to conformational classes would therefore be unreliable in this regime, whereas the ensemble as a whole is still well determined by the full data set.
Indeed, although the resolution achieved here for 1GH9 from synthetic data (about $2\,\mathrm{nm}$) is insufficient for atomistic reconstruction, such a resolution would already be a considerable improvement for experimental single-molecule XFEL data for a structurally \emph{homogeneous} particle of this size.

For the applications presented here, the number of ensemble members (ten in both cases) was chosen ad hoc; choosing this number systematically and optimally remains a challenge. Indeed, a rigorous choice requires computing the Bayesian evidence for different ensemble sizes, for example through replica-exchange sampling. As this would further increase the already considerable computational cost of our method, we did not attempt such an analysis here. Alternatively, the number of members could be increased until the ensemble effectively becomes continuous, which, combined with a suitable prior coupling neighboring members, would allow our approach to directly determine continuous conformational landscapes.

Our synthetic test neglected background scattering, which is unavoidable in actual experiments. However, any background model can be included in the likelihood, as demonstrated here for the coliphage data. Although background scattering reduces the information contained in each image, an accurate error model avoids any bias of the posterior. Indeed, for the determination of single structures, we have previously shown that our Bayesian approach remains applicable in the presence of background scattering~\cite{schultze_bayesian_2024}, and we expect the same to hold for structural ensembles.

Overall, we have established the fact that structural ensembles can be determined even in the extreme Poisson regime. This finding represents an important step towards the determination of conformational ensembles of single proteins, and possibly even quasi molecular movies, from single-molecule XFEL experiments. Combined with steadily improving experimental setups~\cite{rafie-zinedine_coaxial_2025}, such experiments now appear within reach. 

\FloatBarrier

\section{Methods}
\subsection{Likelihood function}
For the synthetic test case we assumed a likelihood function with negligible background scattering, and a perfect detector and a constant pulse intensity $I_0$. In this setting, the likelihood $ P(\vec k_1, \dots, \vec k_n \,|\, \rho)$ of an image $\vec k_1, \dots, \vec k_n$ given an electron density $\rho$ is given by a Poisson distribution for the number of scattered photons $n$ and a product over probability for each observed photon positions, given by the Fourier transform $\hat\rho$ of $\rho$,
\begin{equation}
    P(\vec k_1, \dots, \vec k_n \,|\, \rho) = \int_{\mathrm{SO(3)}} \! \mathrm{Pois}(n \,|\, I_0 p_\mathrm{s}(\vec R, \rho)) \prod_{i=1}^n \frac{\lvert\hat\rho(\vec R\vec k_i)\rvert^2}{p_\mathrm{s}(\vec R, \rho)} \,\,\mathrm{d}\vec R.
\end{equation}
Here, $p_\mathrm{s}$ denotes the overall scattering probability of each incoming photon, obtained by integrating over Ewald's sphere $E$,
\begin{equation}
   p_\mathrm{s}(\rho, \vec R) = \int_E \lvert\hat\rho(\vec R\vec k)\rvert^2 \mathrm d \vec k.
\end{equation}

For the coliphage PR772 data set, we also included an anisotropic background $B(k)$ as well as a detection weight function $D(\vec k)$ to encode the detector geometry, and marginalized over the unknown relative strength $\eta \in [0, 1]$ of the particle scattering compared to the background. Further, because the pulse intensities $I_0$ are random and unknown for each image in this case, we here assumed an uninformative log-uniform prior $P(I_0) \propto 1/I_0$ for the pulse intensities. As $\int_0^\infty \mathrm{Pois}(n \,|\, I_0 p_\mathrm{d})I_0^{-1} \, \mathrm d I_0 = 1/n$ is constant, the likelihood becomes
\begin{equation}\label{eq: likelihood coliphage}
    P(\vec k_1, \dots, \vec k_n \,|\, \rho) \propto \int_{\mathrm{SO(3)}} \int_0^1 \prod_{i=1}^n \frac{\eta \lvert\hat\rho(\vec R\vec k_i)\rvert^2 + B(\vec k)}{p_\mathrm{d}(\vec R, \rho, \eta)} \,\,\mathrm d \eta \,\mathrm{d}\vec R,
\end{equation}
where $p_\mathrm{d}(\vec R, \rho,\eta)$ denotes the overall scattering and detection probability of an incoming photon,
\begin{equation}
   p_\mathrm{d}(\rho, \vec R, \eta) = \int_E D(\vec k)\left(\eta \lvert\hat\rho(\vec R\vec k_i)\rvert^2 + B(\vec k) \right)\mathrm d \vec k.
\end{equation}
Because the photon counts in these images are large, small inaccuracies of the model, for example, in the gain of individual pixels, can strongly bias the resulting densities. We therefore tempered the likelihood of each image for each pose, that is, raised the integrand in eq.~\eqref{eq: likelihood coliphage} to the power of $\beta = 0.1$ before integrating over $\vec R$ and $\eta$. Note that this per-pose tempering is distinct from the global temperature $T$ used for annealing below, which tempers the full likelihood after marginalization.

\subsection{Background modeling}
Because no background-only images were deposited for the coliphage data set, we determined the background $B(\vec k)$ by manually fitting an anisotropic background to the images (Supplementary Fig.~S3). It was modeled as an isotropic radial profile, consisting of a power-law decay, a Gaussian contribution at small $\lVert\vec k\rVert$, and a constant offset, modulated by an anisotropic factor,
\begin{equation}
    B(\vec k) = \left( \frac{c_1}{(k + \varepsilon)^2} + c_2\, e^{-s^2 k} + c_3 \right) \left( a + (1 - a)\, e^{-b\, (\vec k \cdot \vec d)^2} \right),
\end{equation}
with $k = \lVert\vec k\rVert$ and $c_1 = 2.5 \times 10^{-5}\,\mathrm{nm}^{-2}$, $\varepsilon = 0.01\,\mathrm{nm}^{-1}$, $c_2 = 0.2$, $s = 25\,\mathrm{nm}$, and $c_3 = 3 \times 10^{-5}$. Here, $\vec d$ is a unit vector along one of the detector axes, $a = 0.2$ is an isotropic floor, and $b = 10\,\mathrm{nm}^2$ sets the width of the band of increased background perpendicular to $\vec d$.
We have checked through independent runs that the obtained ensembles did not depend strongly on the specific shape of this background, as long as the background level was overall high enough. In contrast, without the inclusion of any background we obtained densities with similar artifacts as observed before~\cite{lundholm_considerations_2018}.

\subsection{Structure representation and prior}\label{sec: structure representation}
Each electron density $\rho_k$ of the ensemble was parameterized in real space as a superposition of $n$ identical Gaussian beads of width $\sigma$ at positions $\vec y_{k1}, \dots, \vec y_{kn}$,
\begin{equation}
    \rho_k(\vec r) = \frac{1}{n \left(\sigma\sqrt{2\pi}\right)^{\mkern-2mu 3}} \sum_{i=1}^n \exp \left(-\frac1{2\sigma^2}\lVert \vec r - \vec y_{ki}\rVert^2\right).
\end{equation}
The widths $\sigma$ of the beads was kept constant, such that only the bead positions $\vec y_{ki}$ were sampled.

The prior consisted of a term for the ensemble weights and a confinement term acting on the bead positions of each ensemble member,
\begin{equation}
    \log P(\rho_{1..K}, w_{1..K}) = (\alpha - 1) \sum_{k=1}^K \log w_k + \sum_{k=1}^K \log P_\mathrm{conf}(\vec y_{k,1..n}) + \mathrm{const}.
\end{equation}
Here, the first term was a symmetric Dirichlet prior with concentration parameter $\alpha$, which for $\alpha > 1$ disfavored ensemble members with vanishing weight. The confinement term combined a flat-top Gaussian for the bead positions with a penalty on beads distant from the remaining density,
\begin{equation}
    \log P_\mathrm{conf}(\vec y_{k,1..n}) = -\sum_{i=1}^n\frac{1}{2\sigma_\mathrm{c}^2} \max\bigl(0, \lVert \vec y_{ki} \rVert - r_0\bigr)^2 - \sum_{i=1}^n \varepsilon_\mathrm{nn} \max\bigl(0, d^{(m)}_{ki} - d_\mathrm{nn}\bigr)^2,
\end{equation}
where $d^{(m)}_{ki}$ denotes the distance of bead $i$ to its $m$-th nearest neighbor within the same ensemble member. The first part is uniform within a sphere of radius $r_0$ around the center of mass and decays as a Gaussian of width $\sigma_\mathrm{c}$ outside. The second part requires each bead to have at least $m$ neighbors within the distance $d_\mathrm{nn}$, which suppressed isolated beads and fragmented densities without imposing any specific connectivity or chain topology. For the synthetic 1GH9 test case, the ensemble weights were fixed to be uniform, and we used $r_0 = 40\,\angstrom$, $\sigma_\mathrm{c} = 5\,\angstrom$, $\varepsilon_\mathrm{nn} = 0.5\,\angstrom^{-2}$, $m = 8$, and $d_\mathrm{nn} = 15\,\angstrom$.

For the coliphage PR772 data set, we omitted the neighbor penalty ($\varepsilon_\mathrm{nn} = 0$) and instead included an ensemble-level prior $\log P_\mathrm{ens}(\vec y_{1..K,1..n})$ that coupled the ensemble members to their mean,
\begin{equation}
    \log P_\mathrm{ens}(\vec y_{1..K,1..n}) = -\frac{1}{2\sigma_\mathrm{ens}^2} \sum_{k=1}^K \sum_{i=1}^n \lVert \vec y_{ki} - \bar{\vec y}_i \rVert^2,
    \qquad
    \bar{\vec y}_i = \frac1K \sum_{k=1}^K \vec y_{ki},
\end{equation}
such that ensemble members deviated from a common structure only where supported by the data. Without this prior, we found that the per-conformer likelihoods of each image differed so strongly that the likelihood gradients for some conformers essentially vanished. We used $\alpha = 5$, $r_0 = 50\,\mathrm{nm}$, $\sigma_\mathrm{c} = 5\,\mathrm{nm}$, and $\sigma_\mathrm{ens} = 3\,\mathrm{nm}$.

For the coliphage PR772 data set, additional prior information was included implicitly to avoid biases due to the incomplete detector coverage, which includes a central hole ($q < 0.04\,\mathrm{nm}^{-1}$) and extends only up to $q \approx 0.72\,\mathrm{nm}^{-1}$ at its corners. In consequence, both the radius of gyration of the conformer is badly determined by the data, as well as any feature with a length scale below $2pi/0.72\,\mathrm{nm}^{-1} \approx 8.7\,\mathrm{nm}$. 
We therefore included two additional terms in the normalization by the total expected photon count $p_\mathrm{d}$ in eq.~\eqref{eq: likelihood coliphage}, replacing it with
\begin{equation}\label{eq: likelihood coliphage prior}
    % change to eq for \tilde p_d
    \tilde p_\mathrm{d}(\vec R, \rho, \eta) = p_\mathrm{d}(\vec R, \rho, \eta) + \lambda_1\, \eta \lvert\hat\rho(\vec 0)\rvert^2 + \lambda_2\eta \displaystyle\int_{E,\, \lVert\vec k\rVert > k_1} \lvert\hat\rho(\vec R\vec k)\rvert^2 \mathrm d\vec k,
\end{equation}
and thereby penalized densities that would scatter a considerable fraction of their intensity into these regions. In effect, the first term acts as a prior towards compact densities, and the latter as a smoothness prior. 
Both terms were implemented by modifying the effective detector geometry $D(\vec k)$ used to compute $p_\mathrm d$. First, the detection weight was set to $1$ within $\lVert\vec k\rVert < k_0 = 0.001\,\angstrom^{-1}$, corresponding to $\lambda_1 = \pi k_0^2$. Second, the detector grid was extended beyond its physical edges up to $q_\mathrm{max} = 0.12\,\angstrom^{-1}$, and a detection weight of $\lambda_2 = 0.1$ was added to all pixels with $\lVert\vec k\rVert > k_1 = 0.065\,\angstrom^{-1}$.

\subsection{Sampling using thermostatted Hamiltonian dynamics.}
The posterior was sampled with Hamiltonian dynamics coupled to the Bussi--Donadio--Parrinello (BDP) stochastic velocity-rescaling thermostat~\cite{bussi_canonical_2007}. All Gaussian bead positions and the logarithmic ensemble weights were combined into one parameter vector $\vec x$ with conjugate momenta $\vec p$ and unit masses, governed by the Hamiltonian
\begin{equation}
  H_T(\vec x, \vec p) = U_T(\vec x) + \tfrac12 \lVert \vec p \rVert^2,
  \qquad
  U_T(\vec x) = -\frac{1}{T}\log P(\mathcal I \mid \rho_{1..K}^{(\vec x)}, w_{1..K}^{(\vec x)}) - \log P(\rho_{1..K}^{(\vec x)}, w_{1..K}^{(\vec x)}),
\end{equation}
where $\rho_{1..K}^{(\vec x)}$ and $w_{1..K}^{(\vec x)}$ denote the ensemble densities and weights defined by the parameter vector $\vec x$. The potential energy $U_T$ is thus the negative log-posterior of this ensemble, with the likelihood tempered by a temperature $T$. 

All runs were initialized at ensembles with uniform weights with each density given by randomly placed beads. To enhance convergence, each run was then started with an annealing phase, in which the temperature was decreased along an exponential schedule,
\begin{equation}
  T_m = T_0^{\,1 - m/M}, \qquad m = 0, \dots, M,
\end{equation}
from an initial temperature $T_0$ to $T_M = 1$ over $M$ integration steps, after which sampling proceeded from the untempered posterior. We used $T_0 = 10$ and $M = 10\,000$ out of a total of $100\,000$ integration steps for the synthetic 1GH9 test case, and $T_0 = 200$ and $M = 50\,000$ out of a total of $100\,000$ integration steps for the coliphage PR772 data set. Only samples drawn at $T = 1$ were used for analysis.

Gradients with respect to the Gaussian bead positions and the ensemble weights were calculated as described before~\cite{schultze_structure_2025}.
The equations of motion were integrated with a velocity Verlet scheme using a time step $\Delta t$,
\begin{equation}
  \vec p \leftarrow \vec p - \tfrac{\Delta t}{2} \nabla U_T(\vec x), \qquad
  \vec x \leftarrow \vec x + \Delta t\, \vec p, \qquad
  \vec p \leftarrow \vec p - \tfrac{\Delta t}{2} \nabla U_T(\vec x).
\end{equation}

After each integration step, the thermostat was applied independently to the bead positions of each ensemble member and to the logarithmic ensemble weights. For each such parameter group, the momenta were rescaled such that the kinetic energy relaxed stochastically towards its canonical distribution~\cite{bussi_canonical_2007}, here with a with relaxation time $\tau = 1$ at a time step $\Delta t = 0.1$.
The center of mass of each density was fixed at zero, and normalization of the ensemble weights was enforced. The corresponding components were projected out of the gradient and the momenta. 

\subsection{Computational aspects and data processing}
The detector was internally represented on a polar grid, uniformly spaced in scattering angle between the inner and outer detector edge, each divided into $n_\varphi = 32$ equal azimuthal sectors. Photons were binned into these pixels, and integrals over the Ewald sphere $E$, such as $p_\mathrm{s}$ and $p_\mathrm{d}$, were evaluated as weighted sums over pixels, weighted by the exact pixel area on $E$. Within each pixel, $\lvert\hat\rho\rvert^2$ was averaged using a $3 \times 3$ point Gauss--Legendre quadrature in polar and azimuthal angle.

The integral over $\mathrm{SO}(3)$ was factorized into the beam direction and the rotation about the beam axis. Beam directions were integrated with a Lebedev quadrature of algebraic degree $23$ ($194$ nodes). Rotations about the beam axis were integrated with a uniform $n_\varphi$-point rule matched to the azimuthal sectors, such that each of these rotations amounts to a cyclic shift of the pixel indices and requires no further evaluation of $\hat\rho$. This yielded $194 \times 32 = 6\,208$ orientations per image and ensemble member. The integral over the signal fraction $\eta$ was evaluated with $N_\eta = 10$ quadrature nodes.

For the test on synthetic data, the images were smoothened by rejection sampling, keeping each photon with a probability given by the Fourier transform of a Gaussian kernel with $\sigma = 2\,\angstrom$. For the test on the coliphage PR772 data, we used the $14\,772$ images classified as hits using the chi-squared metric from the published coliphage PR772 data set~\cite{reddy_coherent_2017}. Images in which any pixel recorded $1\,000$ or more photons were discarded, as the central intensities of these images exceeded the dynamic range of the detector, leaving $N = 12575$ images. 
Pixels within $\lVert\vec k\rVert < 0.04\,\mathrm{nm}^{-1}$ of the beam center as well as a small number of apparently faulty pixels were masked. The integral over the relative signal intensity $\eta$ was implemented by a ten-point Gauss-Legendre quadrature rule.

\subsection{Rotational alignment and conformational sorting.}
The sampled ensembles were rotationally aligned and their members sorted into consistent conformational classes. We denote the $N$ sampled ensembles by $\{\rho_{i1},\dots,\rho_{iK}\}$, $i=1,\dots,N$, with weights $w_{i1},\dots,w_{iK}$, and the Gaussian bead positions of a structure $\rho$ by $\vec y^{\,(\rho)}_{1..n}$. Two structures were compared by the symmetric Chamfer distance of their bead positions,
\begin{equation}
    d(\vec y_{1..n}, \vec y'_{1..m}) = \frac1n \sum_{i=1}^n \min_{j=1}^m\, {\lVert\vec y_i - \vec y_j'\rVert}^2 + \frac1m \sum_{j=1}^m \min_{i=1}^n\, {\lVert\vec y_i - \vec y_j'\rVert}^2 .
\end{equation}

Alignment and sorting were performed jointly by a $k$-medoids scheme with $K$ medoids $M_{1..K}$, which were initialized with the members of a randomly chosen ensemble. In each iteration, every ensemble $i$ was first matched to the medoids by
\begin{equation}
    \bigl(\sigma_i,\;R_{i1},\dots,R_{iK}\bigr)
    = \operatorname*{arg\,min}_{\sigma\in\mathfrak{S}_K,\ R_k\in O(3)}
      \sum_{k=1}^{K} d\bigl(R_k\,\vec y^{\,(i,\sigma(k))}_{1..n},\ \vec y^{\,(M_k)}_{1..m}\bigr).
\end{equation}
To this end, the minimization over $\mathrm{O}(3)$ was carried out numerically for all $K^2$ pairs of ensemble members and medoids, and the optimal permutation $\sigma_i$ was then determined with the Hungarian algorithm on the resulting cost matrix. Each ensemble was subsequently permuted and rotated accordingly,
\begin{equation}
    \rho_{ik} \leftarrow R_{ik}\,\rho_{i\sigma_i(k)},
    \qquad
    w_{ik} \leftarrow w_{i\sigma_i(k)},
\end{equation}
and the medoids were updated,
\begin{equation}
    M_k \;=\; \operatorname*{arg\,min}_{\rho \in \{\rho_{1k},\dots,\rho_{Nk}\}}\;
      \sum_{j=1}^{N} d\bigl(\vec y^{\,(\rho)}_{1..n},\ \vec y^{\,(j,k)}_{1..m}\bigr),
    \qquad k = 1,\dots,K .
\end{equation}
These two steps were alternated until all $\sigma_i$ were the identity. The resulting aligned and sorted ensembles, together with the medoids $M_k$ and mean weights $\bar w_k = \frac1N\sum_{i=1}^N w_{ik}$, were used for all further analysis.

\subsection{Uncertainty and resolution estimate}
The uncertainty of the obtained electron densities was estimated using Fourier shell correlations~\cite{van_heel_fourier_2005},
\begin{equation}
    \mathrm{FSC}(q, \rho_1, \rho_2) = \frac{\int_{\lVert\vec q\rVert = q} \overline{\hat\rho_1(\vec q)}\hat\rho_2(\vec q)\,\mathrm d\vec q}{\sqrt{\int_{\lVert\vec q\rVert = q} |\hat\rho_1(\vec q)|^2\,\mathrm d\vec q}\sqrt{\int_{\lVert\vec q\rVert = q} |\hat\rho_2(\vec q)|^2\,\mathrm d\vec q}}\,,
\end{equation}
where $\rho_1$ and $\rho_2$ are two rotationally aligned densities to be compared, $q$ is the radius of the Fourier shell, and $\hat\rho$ denotes the Fourier transform of $\rho$. FSC curves were calculated for each conformational class $k$ of the aligned and sorted ensembles by comparing all pairs of posterior samples $\rho_{ik}$ and $\rho_{jk}$ and averaging over all $N(N-1)/2$ distinct pairs,
\begin{equation}
    \langle {\mathrm{FSC}}_k(q) \rangle = \frac{2}{N(N-1)} \sum_{i=1}^{N} \sum_{j=i+1}^{N} \mathrm{FSC}(q, \rho_{ik}, \rho_{jk}).
\end{equation}
As all densities within a class were aligned to the same medoid, no further alignment was required.

As an alternative, real-space measure of uncertainty we used Wasserstein distances~\cite{cuturi_sinkhorn_2013}. To this end, each density was represented by a point cloud drawn from it with an equal number of points drawn from each Gaussian component, and the earth-mover's distance $W(\rho, \rho')$ between two densities was computed from the Euclidean distances between their points with the Sinkhorn algorithm, using an entropic regularization of $\varepsilon = 0.1\,\angstrom$. For each conformational class $k$, we summarized the spread of the posterior by the mean pairwise distance
\begin{equation}
    d_k = \frac{2}{N(N-1)} \sum_{i=1}^{N} \sum_{j=i+1}^{N} W(\rho_{ik}, \rho_{jk}).
\end{equation}
The distance between all posterior samples and the reference was instead quantified by the Wasserstein distances between the per-conformer average densities for both.

\subsection{Molecular dynamics trajectories}
To provide ground truth data for the PDB 1GH9 test case, we performed $100\times 10\,\mu\mathrm{s}$ MD simulations using the \textsc{Gromacs 2018} software package~\cite{pronk_gromacs_2013} with the \textsc{Amber99sb‐ildn} force field~\cite{lindorfflarsen_improved_2010} and \textsc{TIP4P-Ew} water model~\cite{horn_development_2004}. 
The starting structure for the simulations was selected from an earlier simulation with the same force field.
It was placed in a triclinic simulation box extending $1.5\,\mathrm{nm}$ beyond the protein in all directions.
Solvent (water) and ions (Na$^+$ and Cl$^-$) were added, establishing a salt concentration of $0.15\,\mathrm{mol}\,\mathrm{l}^{-1}$ and neutralizing the overall system charge.
Energy minimization was performed using \textsc{Gromacs} steepest descent and converged to machine precision (single) after $7\,304$ steps.
Subsequently, two equilibration runs were performed for $0.5\,\mathrm{ns}$ (NVT) and $1\,\mathrm{ns}$ (NPT) at an integration time step of $2\,\mathrm{fs}$.
Then, the $100$ production runs (NPT) were started from the equilibrated system, using an integration time step of $4\,\mathrm{fs}$.
A velocity rescaling thermostat~\cite{bussi_canonical_2007} was applied at $T = 300\,\mathrm{K}$ with temperature coupling constant $\tau_T = 0.1\,\mathrm{ps}$. Isotropic pressure coupling was applied at $p = 1\,\mathrm{bar}$ with pressure coupling constant $\tau_p = 1\,\mathrm{ps}$, using Berendsen pressure coupling~\cite{berendsen_molecular_1984} during equilibration and Parrinello-Rahman pressure coupling~\cite{parrinello_polymorphic_1981} in the production runs. 
All bond lengths were constrained, using the \textsc{Settle} algorithm~\cite{miyamoto_settle_1992} for the solvent and \textsc{Lincs}~\cite{hess_lincs_1997} for the solute, with a \textsc{Lincs} order of $4$ during energy minimization and equilibration and \textsc{Lincs} order of $6$ in the production runs. 
Hydrogen atoms were treated using virtual sites.
Van-der-Waals forces were cut off at $1\,\mathrm{nm}$.
Coulomb forces were calculated using the particle mesh Ewald method (PME)~\cite{darden_particle_1993} with a real-space cutoff of $1\,\mathrm{nm}$, PME order of $4$, and a Fourier grid spacing of $1.2\,\angstrom$. From the MD-trajectories, $6\,125$ frames were selected to serve as the ground truth ensemble. A further $10\,\%$ of these were aligned and sorted into $10$ classes for comparison to the reconstruction.

\section*{Acknowledgments}
This work was supported by the the Deutsche Forschungsgemeinschaft (DFG, German Research Foundation) - CRC 1456/1 - 432680300. The authors gratefully acknowledge the scientific support and HPC resources provided by the Erlangen National High Performance Computing Center (NHR@FAU) of the Friedrich-Alexander-Universität Erlangen-Nürnberg (FAU) under the NHR project j100fa. NHR funding is provided by federal and Bavarian state authorities.

\section*{Data Availability}
We have implemented our method in the Julia programming language~\cite{bezanson_julia_2017,schultze_nphotonsjl_2026}. Source code and data underlying this study are available in Zenodo at \href{http://dx.doi.org/10.5281/zenodo.24158430}{doi:10.5281/zenodo.24158430}. 

% \clearpage 

\printbibliography

\clearpage
\section*{Supplement}
\renewcommand{\thefigure}{S\arabic{figure}}
\setcounter{figure}{0}

\begin{figure}[ht]
    \centering
    \includegraphics[width=0.6\linewidth]{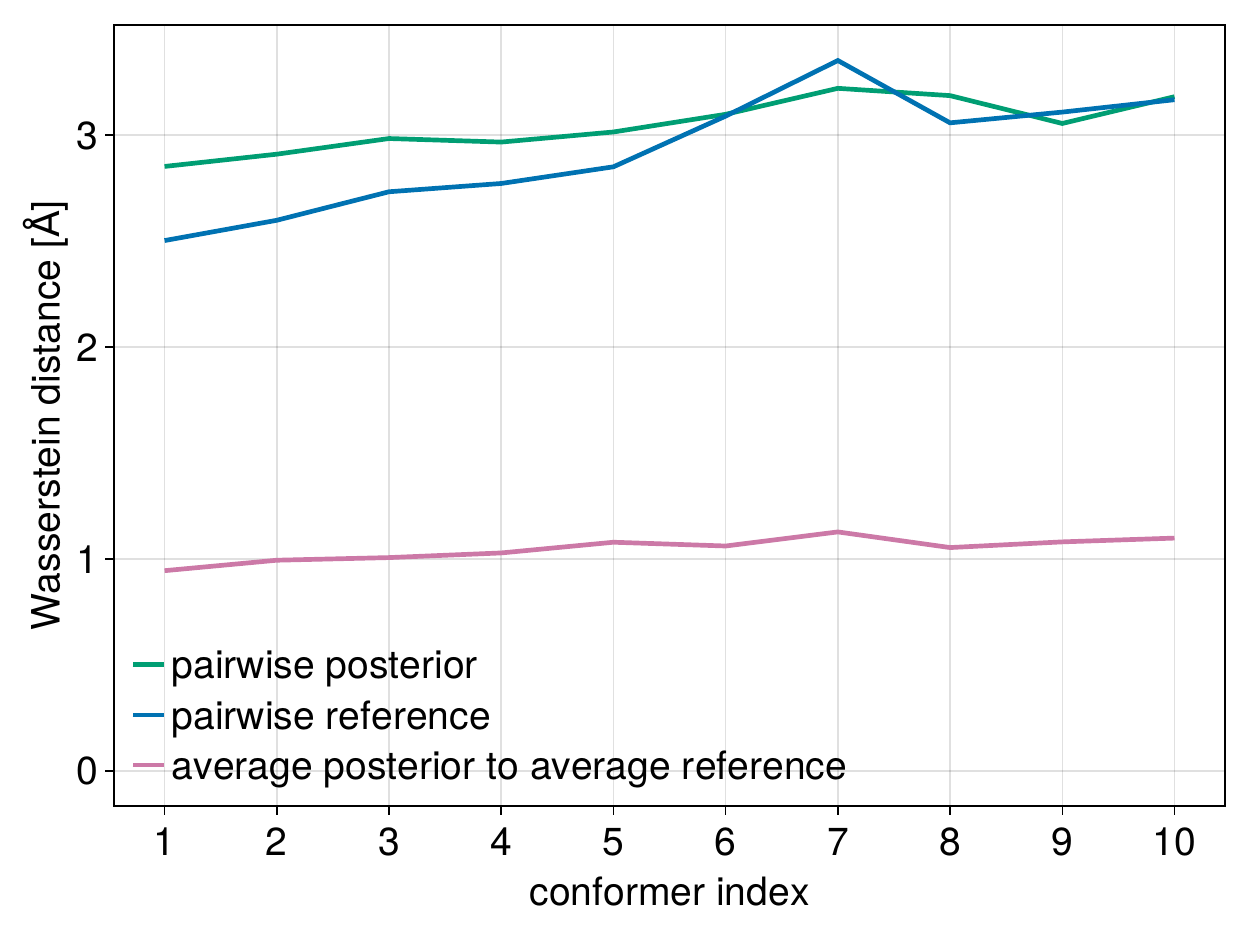}
    \caption{Uncertainty estimate for the 1GH9 test case using optimal transport. Shown are average pairwise internal Wasserstein distances for each conformational class for both posterior and reference ensembles, as well as the Wasserstein distance between the averaged posterior and reference densities.}
    \label{fig: 1gh9_wasserstein}
\end{figure}

\begin{figure}[ht]
    \centering
    \includegraphics[width=1\linewidth]{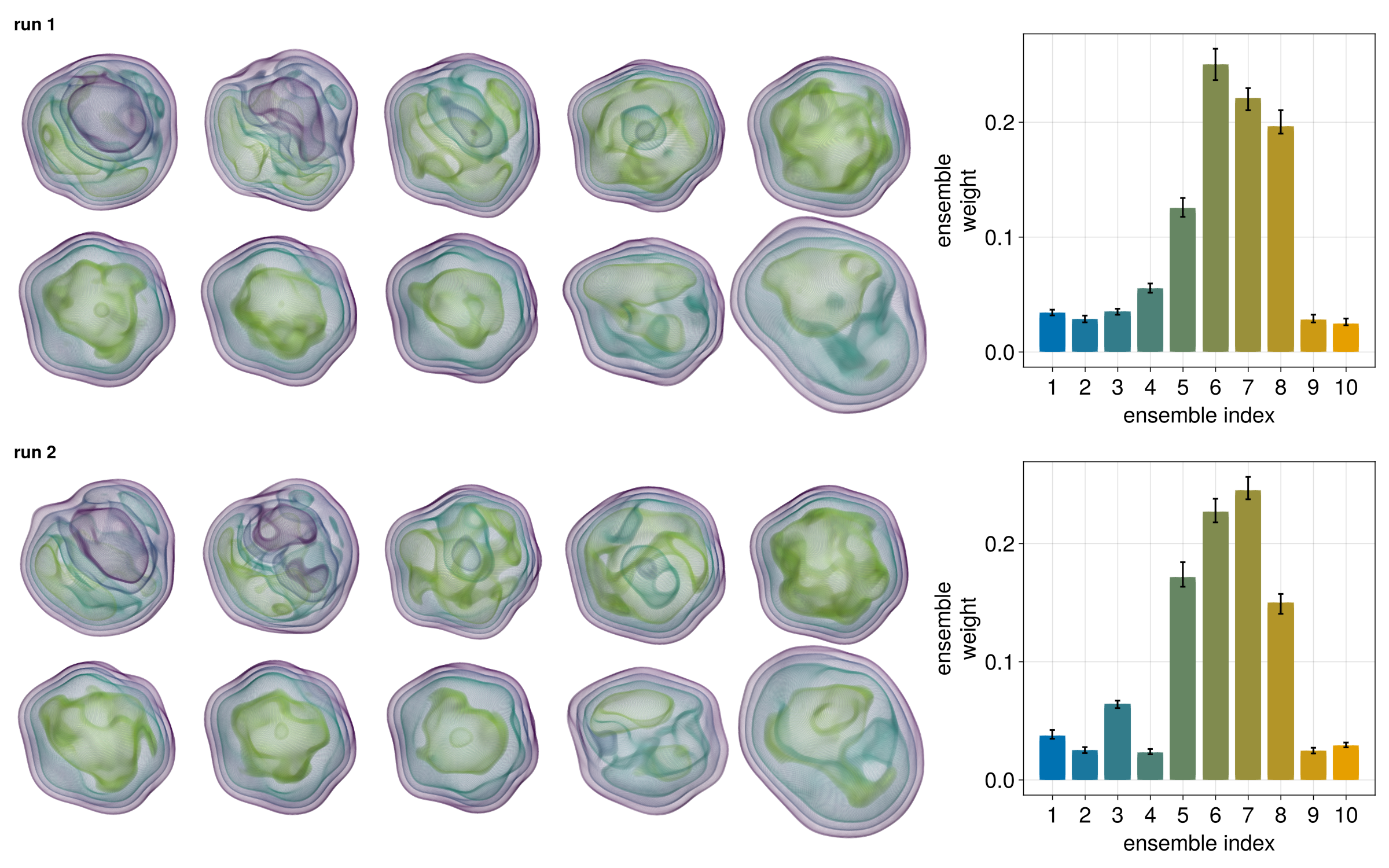}
    \caption{Comparison of densities obtained from two independent sampling runs for the coliphage PR772 application. Shown are the ten average densities (left) as well as their weights (right). Error bars show minimum and maximum weights across each sampling run.}
    \label{fig: coliphage multirun}
\end{figure}

\begin{figure}[ht]
    \centering
    \includegraphics[width=0.6\linewidth]{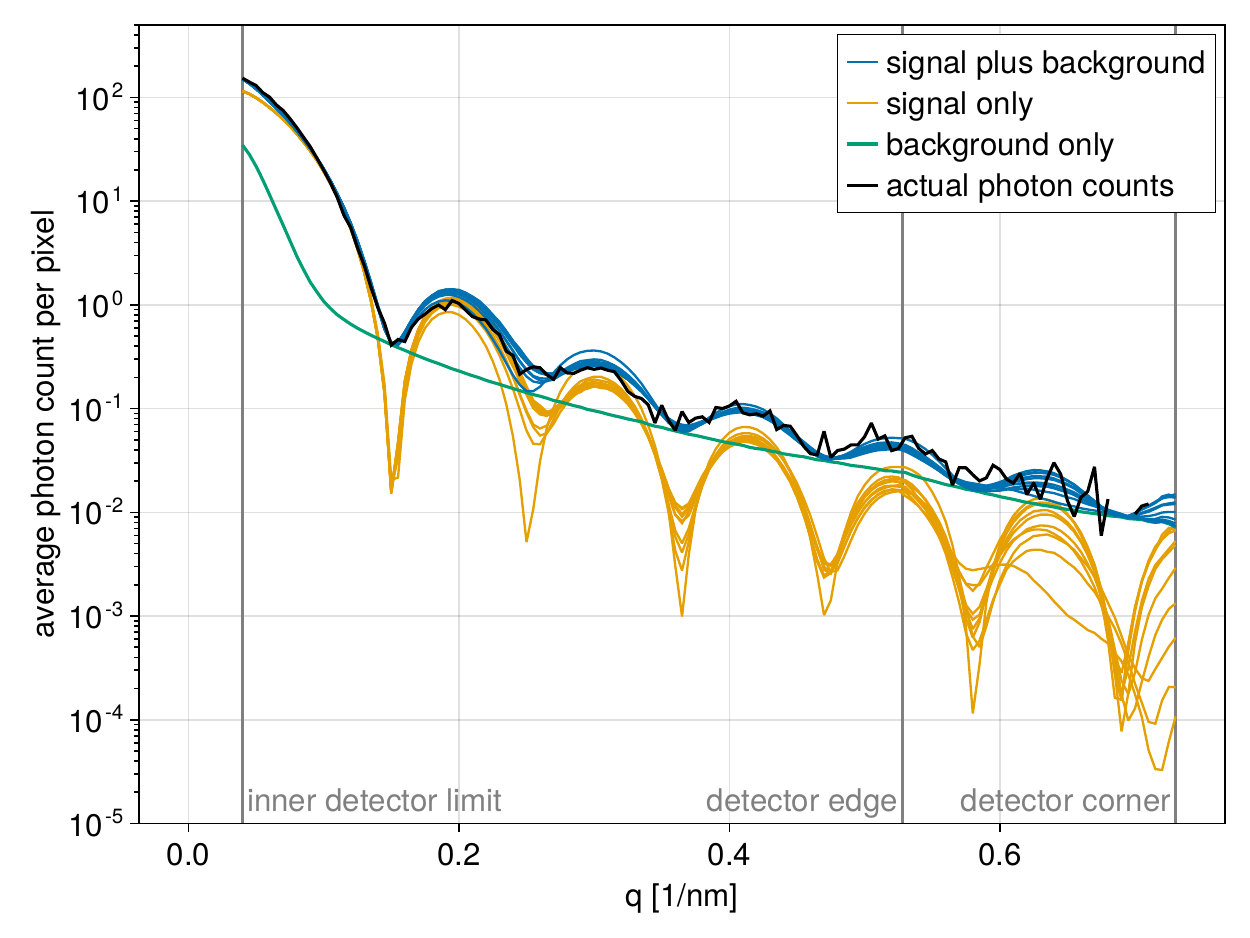}
    \caption{Comparison of experimental photon counts (black) for a single exemplary image (index 2610, hand-selected to feature a single hit corresponding to an icosahedral shape). The signal (orange) was calculated for a uniformly filled icosahedron of radius $34\,\mathrm{nm}$. For signal plus background (blue) and signal only, each line represents a different random orientation.}
    \label{fig: coliphage signalandbackground}
\end{figure}

\end{document}